\documentclass[letterpaper, 10 pt, conference]{ieeeconf}  

\IEEEoverridecommandlockouts                              

\usepackage{amsmath} 
\usepackage{graphicx} 
\usepackage{pythonhighlight}
\usepackage[T1]{fontenc}
\usepackage{amsfonts}
\usepackage{acro}
\usepackage{siunitx}
\usepackage{hyperref}
\usepackage{dirtytalk}
\usepackage{makecell}
\usepackage[normalem]{ulem}
\usepackage[hang,flushmargin]{footmisc}

\title{\LARGE \bf
MARLadona - Towards Cooperative Team Play Using Multi-Agent Reinforcement Learning
}

\renewcommand{\baselinestretch}{0.987}
\newcommand{\edited}[2]
{{#2}}
\author{Zichong Li, Filip Bjelonic, Victor Klemm, and Marco Hutter  
\thanks{This paper has been accepted for publication at the IEEE International Conference on Robotics and Automation (ICRA) 2025. \newline
All authors are with the Robotic Systems Lab, ETH Z\"urich, 8092 Z\"urich, Switzerland.}
}

\DeclareAcronym{rl}{short=RL,  long=reinforcement learning}
\DeclareAcronym{marl}{short=MARL, long=multi-agent reinforcement learning}
\DeclareAcronym{mas}{short=MAS, long=multi-agent system}
\DeclareAcronym{gee}{short= GEE, long=global entity encoder}
\DeclareAcronym{soa}{short= SOTA, long=state-of-the-art}
\DeclareAcronym{rc}{short= RC, long=RoboCup}
\DeclareAcronym{mlp}{short= MLP, long=multilayer perceptron}
\DeclareAcronym{mdp}{short= MDP, long=markov decision process}
\DeclareAcronym{ctde}{short= CTDE, long=centralized training and decentralized execution}
\DeclareAcronym{mg}{short= MG, long=Markov game}
\DeclareAcronym{pomg}{short= POMG, long=partial observable Markov game}
\DeclareAcronym{decpomdp}{short= Dec-POMDP, long=decentralized partially observable Markov decision process}
\DeclareAcronym{decpomg}{short= Dec-POMG, long=decentralized partially observable Markov game}

\begin{document}

\maketitle
\thispagestyle{empty}
\pagestyle{empty}
\urlstyle{same}
\setlength{\textfloatsep}{7pt}
\setlength{\dbltextfloatsep}{7pt}
\setlength{\floatsep}{7pt} 
\setlength{\intextsep}{7pt}
\begin{abstract}
  Robot soccer, in its full complexity, poses an unsolved research challenge. Current solutions heavily rely on engineered heuristic strategies, which lack robustness and adaptability. Deep reinforcement learning has gained significant traction in various complex robotics tasks such as locomotion, manipulation, and competitive games (e.g., AlphaZero, OpenAI Five), making it a promising solution to the robot soccer problem.  
  This paper introduces MARLadona. A decentralized \ac{marl} training pipeline capable of producing agents with sophisticated team play behavior, bridging the shortcomings of heuristic methods. Furthermore, we created an open-source multi-agent soccer environment\edited{based on Isaac Gym}. Utilizing our \ac{marl} framework and a modified \edited{\iacl{gee}}{\acf{gee}} as our core architecture, our approach achieves a \SI{66.8}{\percent} win rate against HELIOS agent, which employs a state-of-the-art heuristic strategy. In addition, we provided an in-depth analysis of the policy behavior and interpreted the agent's intention using the critic network.
\end{abstract}
\section{INTRODUCTION}
        \label{sec:introduction}
        Recent advancements in deep \ac{rl} have made it possible to solve some of the most challenging robotic tasks such as object manipulation \cite{wang2019stable}, locomotion \cite{lee2020learning}, and high-level navigation tasks \cite{lee2024learning}. However, many real-world problems require cooperation among multiple robots, requiring \ac{rl} agents to learn optimal behavior while coexisting with other agents, with either conflicting or common objectives requiring negotiation and communication. As a result, this intricate interplay of strategies introduces additional complexity, making \ac{mas} more challenging than single-agent systems. 

As one of the most popular \ac{mas}, robot soccer has been the grand goal of robotic and AI research. Competitions such as \ac{rc} \cite{kitano1998robocup} and IEEE Very Small Size Soccer \cite{martins2021rsoccer} have inspired generations of researchers to push the \ac{soa} in robotic hardware and algorithms. Traditionally, the high-level strategies of soccer agents are often hierarchically separated from the rest and tackled with sophisticated decision trees crafted by human experts. But robot soccer as a pure learning problem has gained significant attention in recent years \cite{liu2022motor, haarnoja2024learning}. Existing literature in soccer \ac{marl} can be roughly categorized into creating environments and proposing training approaches. 
\begin{figure}[t]
    \centering
    \includegraphics[width=0.42\textwidth]{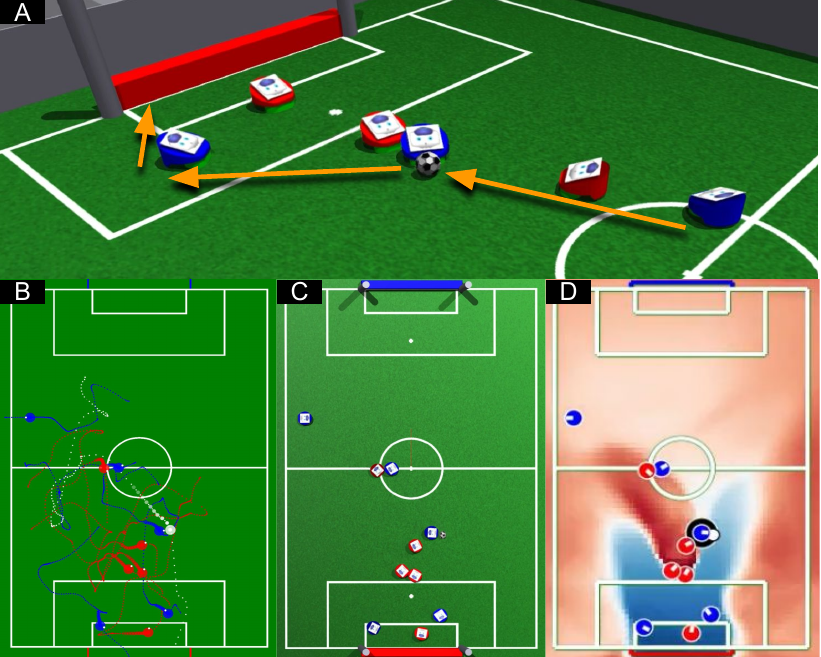}
    \caption{An illustration of our \ac{marl} environment (A) and various top-down views of a 5v5 game (B-D). (B) The trajectory visualizer depicts the general game dynamic. (C) The corresponding default top-down view. (D) The corresponding ball position critic value heat map. }
    \label{fig:intro}
\end{figure}
\begin{figure*}
    \centering
    \includegraphics[width=0.88\textwidth]{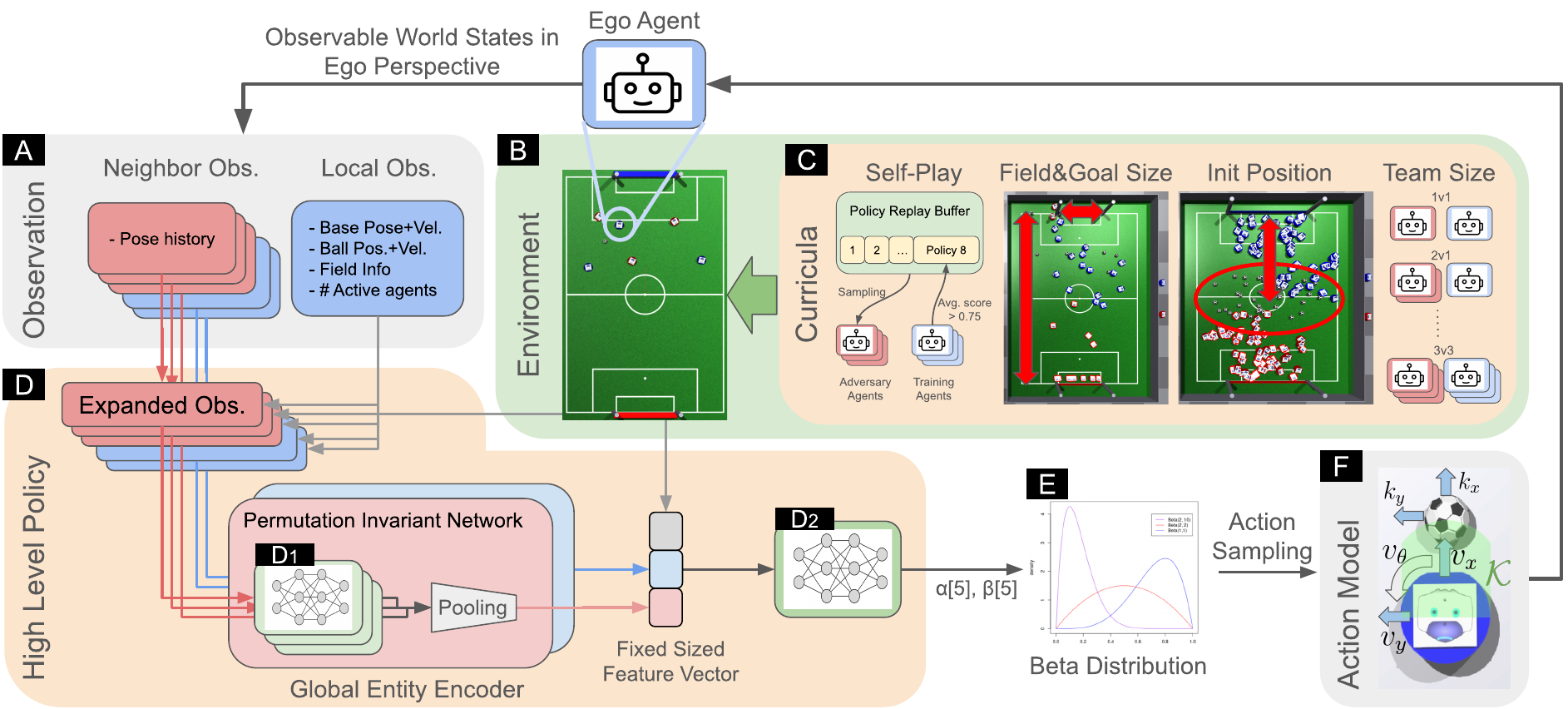}
    \caption{An overview of our system. (A) The ego perspective observation from the opponents (red), teammates (blue), and local observation. (B) The soccer environment. (C) Various curricula we adopted during training. (D) The architecture of our policy network. (D1) Encoders with shared weights. (D2) Policy network. (E) The distribution we used for action sampling. (F) The action model of our soccer agent.}
    \label{fig:overview}
\end{figure*}

\subsection{Related Work}
Several soccer environments have already been proposed by literature such as Google research football \cite{kurach2020google}, MARL2DSoccer \cite{smit2023scaling}, rSoccer \cite{martins2021rsoccer}, Humanoid football in MoJoCo \cite{liu2019emergent} and the \ac{rc} Simulation2D server \cite{kitano1998robocup}. However, besides lacking GPU support, many existing \ac{marl} soccer environments are kinematic simulators \cite{smit2023scaling, martins2021rsoccer, kitano1998robocup}. The lack of physics creates an additional sim-to-real gap, making transferring the learned strategies to real robotic platforms harder. On the other hand, \ac{marl} environments like \cite{kurach2020google} are great for comparing novel algorithms by providing a standardized framework. Still, its game-inspired actions, i.e., abstracted action like \say{long pass}, make it even less applicable. Most \ac{rc} teams also have their own customized multi-agent soccer simulators \cite{bhuman, aggarwaltech, ratzel2023robocup} tailored and fine-tuned for their specific robotic platforms.  

While existing approaches for \ac{marl} soccer often tackle the multi-agent strategy as a stand-alone \ac{marl} problem \cite{smit2023scaling, an2024solving, catacora2019cooperative}. Recent works from DeepMind \cite{liu2022motor, haarnoja2024learning} have demonstrated an approach that solves the full soccer problem end-to-end in a 3D environment. Both of these works utilized a combination of imitation learning and individual skill training to obtain low to mid-level motor skills. Then, they trained with \ac{marl} in a self-play setting to get the final policy. While emerging team play was observed in \cite{liu2022motor}, they only demonstrated within a 3v3 game setting.
For games with more agents, it is more common to simplify the soccer problem into 2D and train high-level policies \cite{an2024solving, catacora2019cooperative, smit2023scaling, martins2021rsoccer,zhong2023development} which assume the existence of motion interfaces such as walking velocity and kicking commands. Existing work commonly trains decentralized policies \cite{smit2023scaling, lin2023tizero} to avoid the exploding state space with growing agent number. For instance, Smit et al. \cite{smit2023scaling} used an attention mechanism and utilized their customized MARL2DSoccer environment for simulation. Lin et al. \cite{lin2023tizero} used Google research football directly and presented the TiZero, which utilized a combination of observation encoder and LSTM mechanism with a multi-stage self-play curriculum. They achieved \ac{soa} performance, but their game-inspired actions make their policy incompatible with actual robotic systems. 
      
For many \ac{mas}, permutation invariant with respect to neighboring agents is a desirable property. An indirect way to achieve this is to augment the data with the permutated state, similar to learning symmetry \cite{mittal2024symmetry}. However, it is more desirable to have it directly as a network property. 
Various forms of permutation invariant networks have been proposed in the literature, e.g., Attention mechanism \cite{vaswani2017attention}, PointNet \cite{qi2017pointnet}, deep sets \cite{zaheer2017deep} etc. 
Based on the PointNet, An et al. \cite{an2024solving} proposed the \ac{gee} in which set inputs of the same type, called entities, are passed through a shared \ac{mlp} allowing a list of local feature encodings to be obtained. The max pooling is then conducted over the set of local feature encodings and the resulting global entity feature, which is permutation invariant for set inputs. 

\subsection{Contributions}
While existing training approaches often demonstrate improved quantitative results against some heuristics \cite{smit2023scaling, martins2021rsoccer, zhong2023development}, none of them, to our best knowledge, demonstrated quality team play in \ac{marl} soccer for more than 3v3 games.
To address this open research problem, we introduce MARLadona, a novel framework for training high-level multi-agent policies using end-to-end \ac{rl}. 

Our key contributions include:
\begin{itemize}
        \item A new customizable open-source \edited{2D}{}\ac{marl} soccer environment \edited{based on Isaac Gym}{based on Isaac Lab}\footnote{\edited{}{\url{https://github.com/leggedrobotics/marladona-isaac-lab} \\ Note that the original work in this paper was based on Isaac Gym.}}.
        \item Introduced an improved \ac{gee} for training policies that achieve effective team play for games up to 11v11.
        \item Provided a comprehensive analysis of the policy behavior and benchmarked its performance against HELIOS\footnote{For clarity, we use HELIOS to refer to \cite{akiyama2014helios}}, a \ac{soa} scripted soccer agent. 
\end{itemize}

\section{METHOD}
        \label{sec:method}
        \subsection{Overview}
\label{sec:overview}
Our work aims to train a generic decentralized soccer policy that demonstrates effective and dynamic team play. To achieve this, we designed a multi-agent soccer environment on top of Isaac Gym to leverage its realistic physics and GPU acceleration and utilized this framework in combination with a novel approach (Fig. \ref{fig:overview}) to train the soccer policy. We address the sparse reward problem of soccer \ac{marl} with different curricula (C) and adopted a modified version of the \ac{gee} architecture \cite{an2024solving} as our actor (D) to effectively handle permutation and number changes among both teammates and opponents. Moreover, our approach aims to include only the minimum for observation states and rewards. By using the PPO algorithm \cite{ppo}, we trained our lightweight network on a consumer desktop (single RTX 2060 GPU) and achieved quality team play within just a few hours of training.         

\subsection{Markov Game}
\label{sec:theory}
Different from traditional single-agent \ac{rl} problems, \ac{marl} the problem is often modeled as a \ac{mg} \cite{littman1994markov} which is a generalization of the \ac{mdp} to multiple agents. A \ac{mg} can be formally defined as a tuple  
    $\langle \mathcal{I}, \mathcal{S}, \mathcal{A}, \mathcal{P}, \mathcal{R}, \gamma \rangle$, where $\mathcal{I}$ denotes the set of agents, $\mathcal{S}$ is the environment state, $\mathcal{A} =\{ a_{i} \}_{i=1}^N $ is the joint action of all agents, $\mathcal{P}: \mathcal{S} \times \mathcal{A} \rightarrow \mathcal{S}$ the transition probability, $\mathcal{R} = \{ r_{i} \}_{i=1}^N$ with $ r_i : \mathcal{S} \times \mathcal{A} \times \mathcal{S} \rightarrow \mathbb{R}$ is the reward function of agent $i$ and $\gamma$ as the discount factor. 
At each time step $t$, the state $\textbf{s} \in \mathcal{S}$ evolves according to the transition probability $\mathcal{P}$ and the joint action $\textbf{a} \in \mathcal{A}$. 
Each agent $i$ then receives their respective reward $r_i$ from the environment as feedback. 

Often in \ac{marl}, the global environment states are not fully observable, leading to the \ac{pomg} \cite{gronauer2022multi}. The additional $\mathcal{O}$ is used to denote the observable states of the environment. In case all agents share the same reward function, then a cooperative \ac{pomg} can be considered as a \ac{decpomdp} \cite{hansen2004dynamic}. 

For this paper, we adopted the \ac{ctde} paradigm due to its scalability and robustness to varying agent numbers and access to privileged information.
In \ac{ctde}, each agent has its policy $\pi_i: \omega_i(\mathcal{O}) \rightarrow a_i$, which maps its local observation to its action $a_i$. $\omega_i$ is the observation function of agent $i$. Denoting the additional set of observation function as $\varOmega = \{ \omega_{i} \}_{i=1}^N$, our final problem can be modeled as a \ac{decpomg} defined by the tuple:
\begin{equation}
    \langle \mathcal{I}, \mathcal{S}, \mathcal{A}, \mathcal{O}, \varOmega , \mathcal{P}, \mathcal{R}, \gamma \rangle 
\end{equation}

\subsection{Environments and Agent Modeling}
\label{sec:environment}
Our \ac{marl} soccer environment (Fig. \ref{fig:overview} B) is built on top of Isaac Gym \cite{makoviychuk2021isaac}, allowing us to simulate thousands of games in parallel. Each environment contains a ball and two teams with a random number of agents, and all actors are uniformly spawned within their designated area. A game terminates when a goal is scored or after 30 seconds, which is empirically chosen to give the agents enough time to score. The ball respawns above the borderline upon leaving the playing field, and a physical wall is placed at a short distance around the playing field to prevent escaping agents. Furthermore, the environment registers events such as passes, ownership\footnote{An agent is assigned with ball ownership if it is within \SI{0.25}{\m}, the closest from his team, and no opponent fulfills the first condition.} losses, etc. Only the blue agents are used for training, while the red agents are controlled by bots or old policies, depending on the scenario. For clarity, we will refer to the blue agent as \textbf{trainee} and the red agent as \textbf{adversary}. 

A floating base with a cylinder as its collision body is used for the agent modeling. At each simulation step, each agent receives an action command of the form ($v_x, v_y, v_{\theta}, k_x, k_y$) (Fig. \ref{fig:overview} F). All actions represent velocity commands\footnote{Velocity commands as it is widely used to interface with robotic platforms.} which is then tracked by a PD controller outputting a corresponding force and torque for the simulator. The $x$ and $y$ component of the translation commands (base: $v_x, v_y$) and (kick: $k_x, k_y$) are remapped using Eq. \ref{eq:remap} to map the rectangular control space into the unit disk. Note, the agent's kick commands are only in effect when the ball is within its kickable area $\mathcal{K}$.
\begin{equation}
    \label{eq:remap}
    x_{new} = x \cdot {\sqrt{1 - \frac{y^2}{2}}} \quad, \quad y_{new} = y \cdot {\sqrt{1 - \frac{x^2}{2}}}
\end{equation}
\begin{figure}
    \centering
    \includegraphics[width=0.45\textwidth]{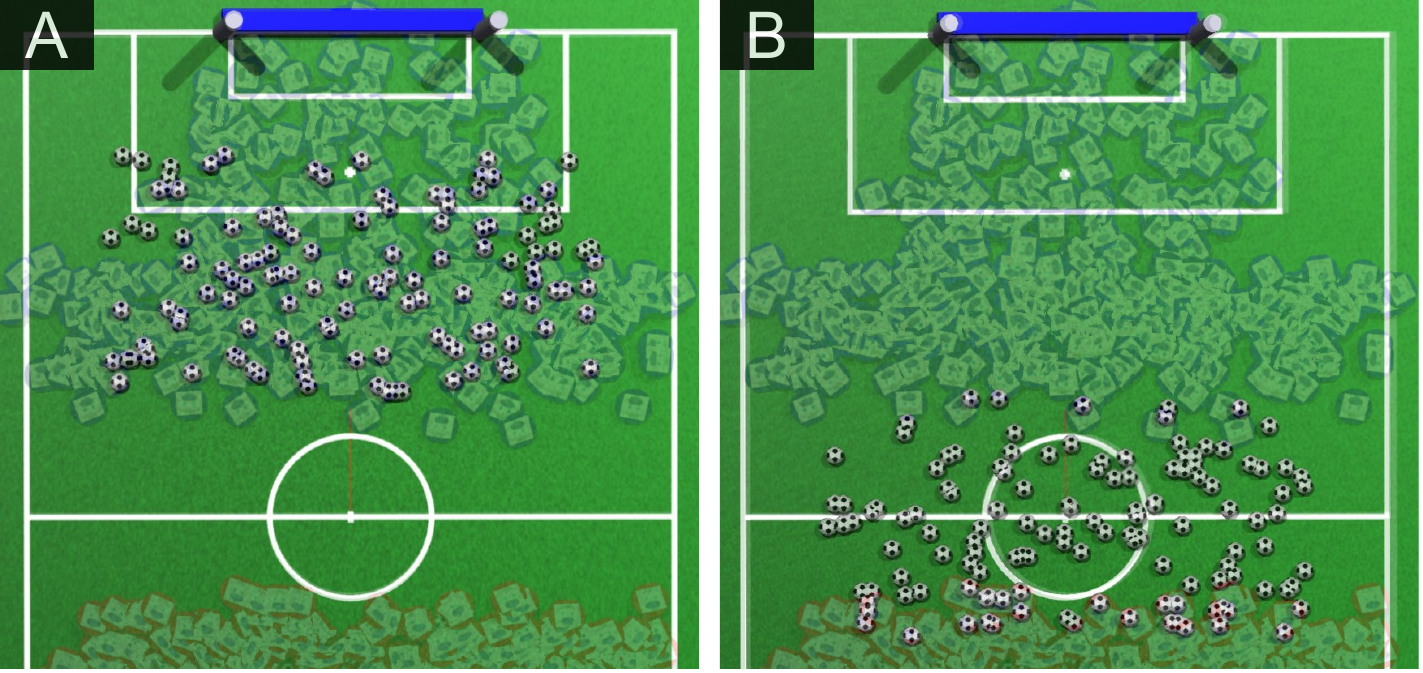}
    \caption{Initial position curriculum depended on the current policy performance. The ball's initial distribution is adjusted toward the blue side for lower levels to enhance trainees' chances of gaining ball procession. The agent's initial distribution, on the other hand, is kept constant. }
    \label{fig:init_pos}
\end{figure}

\begin{figure*}
    \centering
    \includegraphics[width=0.86\textwidth]{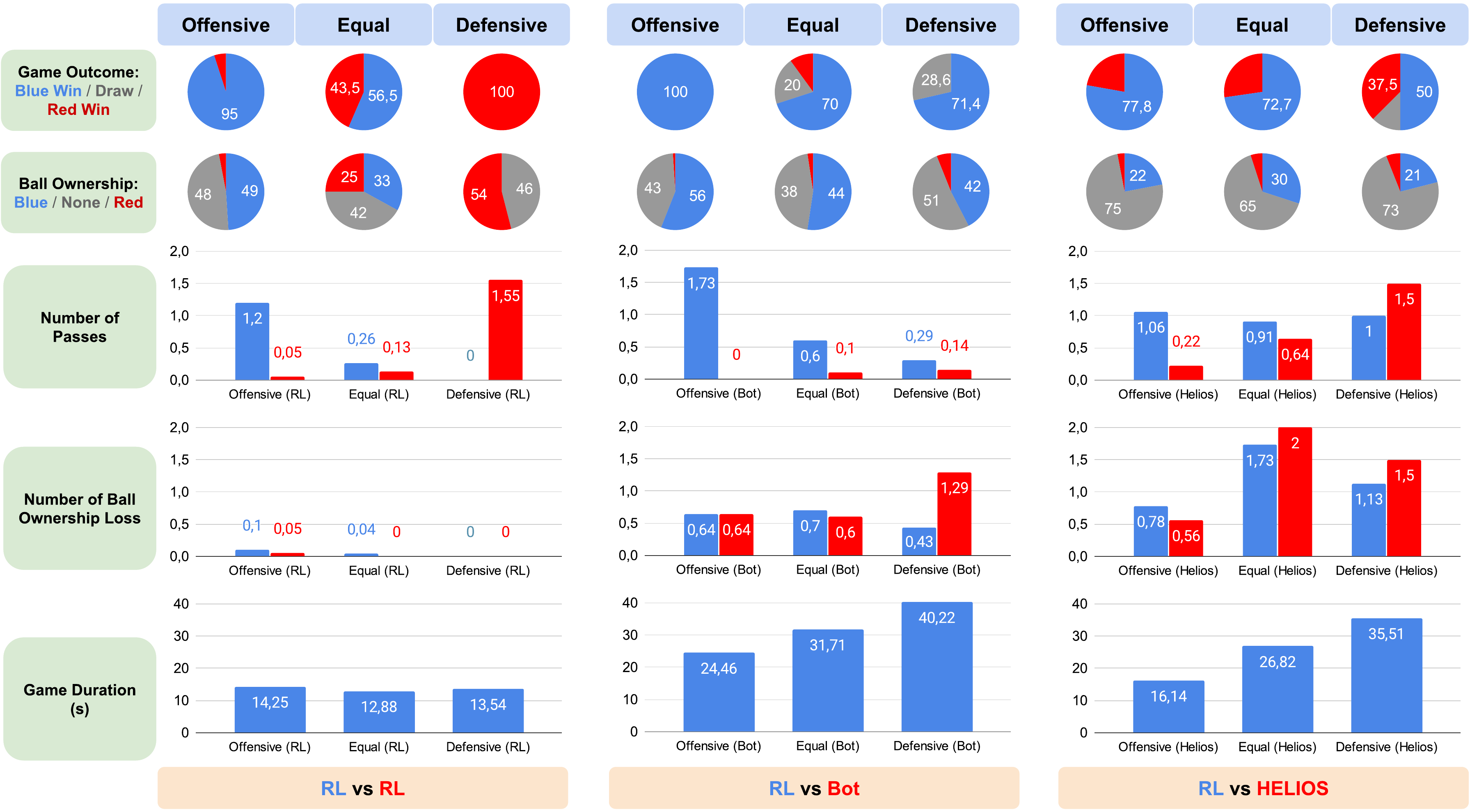}
    \caption{An overview of our evaluation results conducted for a 3v3 game for three different scenarios (Offensive, Equal, Defensive) against three different adversaries (RL, Bot, HELIOS). The collected average statistics (game outcome (\si{\percent}), team ball ownership (\si{\percent}), the number of successful passes and ball ownership losses, and game duration) are depicted in different rows. Our trainee policy (Blue) achieved clear dominance against all adversaries (besides itself) in all scenarios. The trainees won \SI{66.8}{\percent} (averaged over all three scenarios) of all games against HELIOS. }
    \label{fig:eval_results}
\end{figure*}

\subsection{Curricula}
To reduce the initial exploration complexity and accelerate the learning process, we included various types of curriculum during training (Fig. \ref{fig:overview} C). Inspired by terrain curriculum \cite{leggedgym}, we added different field and goal size levels to increase the initial scoring chance and adjusted the environment level depending on the outcome of the previous game. 

Self-play is another approach commonly applied in competitive \ac{rl} problems  \cite{silver2018general,haarnoja2024learning, kurach2020google, smit2023scaling, brandao2022multiagent}. It allows our trainees to adapt to different adversaries and incentivizes them to learn more generalized strategies. Various forms of implementation can be found in the literature, such as League training, Polyak averaging \cite{smit2023scaling}, or sampling from past policies \cite{haarnoja2024learning}. We adopted a replay mechanism that buffers up to 8 policies. During training, the current trainee policy is added to the adversary buffer when it achieves an \SI{75}{\percent} average win rate against all of its adversaries. The adversaries then sample from this buffer and run the policy in inference mode. During sampling, we apply the same policy per team, as cooperation quality might deteriorate due to strategy mix-ups. 

Unlike many other \ac{marl} problems, the initial position of the agents and ball is critical as being closer to the ball directly implies ownership, which is a significant advantage in soccer. Therefore, we adopted different curriculum levels to the initial position as well. In our approach, agents are unaffected by the curriculum level and just spawned uniformly within their designated area with direction sampled from $(\pi, -\pi]$. The ball is biased deeper into the blue side for lower curriculum levels to increase ball ownership probability for the trainee team (Fig. \ref{fig:init_pos}). 

Finally, we also want the trainees to learn and adapt their strategies accordingly, depending on the team sizes. Since training directly with the maximum agent number can be counterproductive, we implemented a resampling mechanism that allows the environment to dynamically re-configuration its team composition before an episode begins. This enables us to adapt the maximum number of players according to the policy performance. Due to computational constraints, we limited our training to a maximum of 3v3 games.  

\subsection{Observations and Rewards}
\label{sec:observation}
\begin{table}
    \caption{Observation Overview. }
    \label{tab:observation}
    \begin{tabular}{ l r r r}    
            Observation name & Noise actor/critic & Dimensions & Options \\ \hline
            \multicolumn{4}{l}{Local Observations} \\ \hline
            1) Base pose & 0.002 / 0 & 4 & W, E, N\\ 
            2) Base velocity & 0.005 / 0 & 3 & W \\ 
            3) Ball position & 0.002 / 0 & 2 & W, N \\ 
            4) Ball velocity & 0.005 / 0 & 2 & W\\
            5) Field info & 0 / 0 & 5 & N \\ 
            6) $\#$ active agents & 0 / 0 & 2 & \\ \hline
            \multicolumn{4}{l}{Neighbor Observations} \\ \hline
            7) Teammate poses & 0.002 / 0 & 4 $\cdot H \cdot N_{T}$ & W, E, N\\
            8) Opponent poses & 0.002 / 0 & 4 $\cdot H \cdot N_{O}$ & W, E, N\\
    \end{tabular}
\end{table}

An overview of the observation is summarized in Tab. \ref{tab:observation}. Rows 1-6 are local observations, and rows 7 and 8 show the neighbor observations. Both actor and critic use the same observation states but slightly different noise configurations. 
Note that all observation types are transformed into the perspective of the ego agent. The option column provides additional configuration details. (W) indicates the observation is in the world frame, (E) denotes the rotation component $\theta$ is expanded into $\sin{(\theta)}$ and $\cos{(\theta)}$. (N) means we normalize the observation with the field length and width. And $H$ shows the number of included past time steps. 
The field info contains auxiliary information about the x and y position of the field line border and goal width. For the neighbor observations, the observation function $\varOmega$ is implemented as a filter mask that passes the $N_{max}$ closest agents' state, which is kept at three during training. For testing with higher agent numbers, e.g., 5v5 and above, we use $N_{max} = 5$ and also limit the value of observation 6 in Tab. \ref{tab:observation} to three active agents, avoiding scalability issues. 

Tab. \ref{tab:reward} provides an overview of our rewards. Rows 1-3 are our main sparse rewards, and rows 4-6 show the dense rewards that get permanently removed the first time the trainee policy reaches \SI{75}{\percent} win rate against its adversaries. The dense rewards are only added to provide initial guidance by incentivizing the trainees to approach the ball if neither the ego agent nor his teammates have ball ownership (reward 4), bring the ball closer to the goal (reward 5), and keep the ball within its kickable area $\mathcal{K}$ (reward 6). The shared column indicates which of them is shared with the whole team. 

\begin{figure*}
        \centering
        \includegraphics[width=0.86\textwidth]{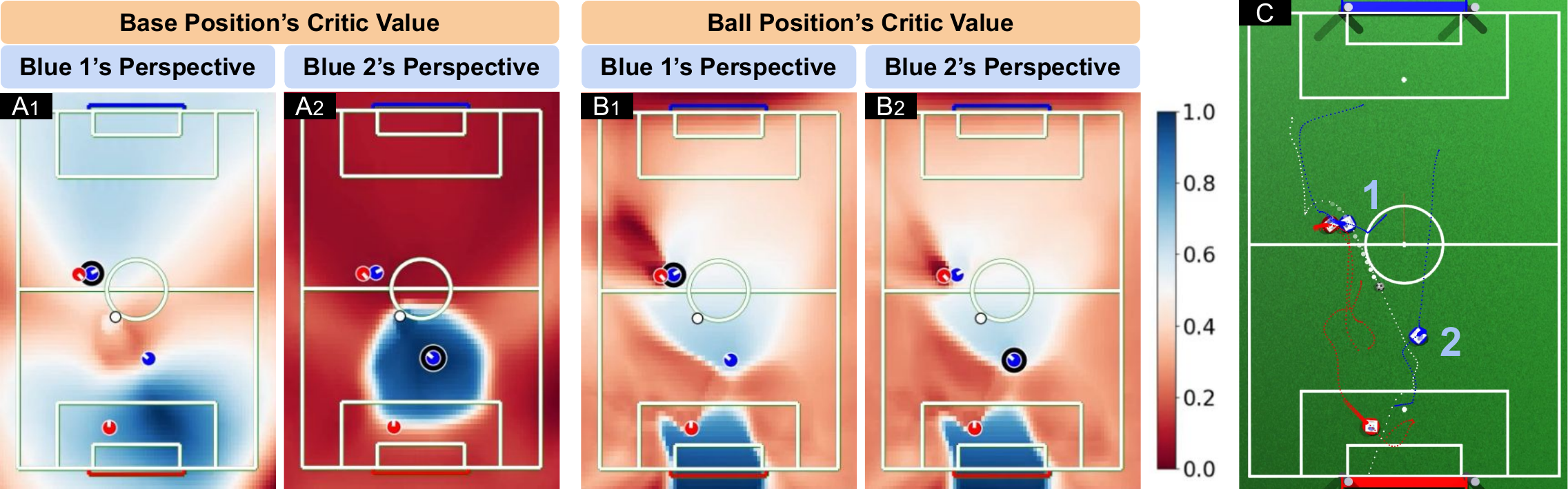}
        \caption{An illustration of the critic values from a 2v2 \edited{}{(RL vs RL)} game as a heat map (Res. 80 $\times$ 80).
      The plots are acquired by varying the base positions of the trainees (A1, A2) and the ball position (B1, B2) over the whole field while keeping the other observations fixed. The position of the trainees (blue), adversaries (red), and ball (white) are represented by their respective colored dots, and the large black circle indicates which of the trainees the heat map belongs to. Furthermore, C provides the corresponding default top-down view overlaid with motion trajectories to provide additional information about the current game dynamic. The blue areas on these heat maps indicate where the agents want themselves and the ball to be.}
        \label{fig:value_function}
\end{figure*}

\begin{table}
    \caption{Reward Overview. }
    \label{tab:reward}
    \begin{tabular}{ l l r r}    
        Reward name & Description & Scale & Shared  \\ \hline
        \multicolumn{4}{l}{Sparse Rewards} \\ \hline
        1) Score & +1 if wins, -1 if loses & 100 & True \\ 
        2) Ball outside field & -1 if ball leaves the field. & 1 & True \\ 
        3) Collision & \makecell{-1 if collision (except ball)} & 1 & \\ \hline
        \multicolumn{1}{l}{Dense Rewards} & \multicolumn{3}{l}{(Removed later)} \\ \hline
        4) Ball2goal velocity & $\lVert v_{ball2goal} \rVert$ & 2 & True \\
        5) Base2ball velocity & $\lVert v_{base2goal} \rVert$ if ball is far & 0.5 &  \\ 
        6) Ball direction & $\exp(-(\theta_{base2ball}/0.4)^2)$ &0.025 &  \\ 
    \end{tabular}
\end{table}

\subsection{Network Architecture}
\label{sec:architecture}
\begin{table}
    \caption{Network Configuration. }
    \label{tab:network}
    \begin{tabular}{ l | r r r r}    
        & Input & Hidden & Output & Activation \\ \hline
        Encoder network & 27 & 64, 32 & 16 & ELU \\ 
        Policy network & 35 & 128, 128, 128 & 10 & ELU \\ 
    \end{tabular}
\end{table}
\ac{mas} often requires agents to have invariant strategies with respect to neighboring agents while dynamically adapting to changes in agent number. We achieved this using a similar architecture as the \ac{gee} \cite{an2024solving} (Fig. \ref{fig:overview} D). 
One key difference to the original \ac{gee} architecture is the expanded observation formed by concatenating each neighboring agent's observation with the local observation. This expanded observation is then forward fed into the encoder network (D1), allowing the shared encoder to extract information with additional context of the ego agent. The most prominent feature of the resulting local entity feature is max pooled across each agent type. The final input for the policy network (D2) can be obtained by merging the fixed-size global entity feature of the teammates and opponents, which can be concatenated with the local observation. The layer size configuration of both the encoder and policy network are summarized in Tab. \ref{tab:network}. Similar to \cite{an2024solving}, we use beta distribution to adhere to the constraint of the velocity command, which requires the policy net to predict a ($\alpha, \beta$) value pair per action.

\section{EXPERIMENTAL SETUP}
        \label{sec:experiment_setup}
        \subsection{Evaluation Methodology}
\label{sec:methodology}
We validated our trainee policy against three adversaries using three scenarios in a 3v3 setting. Over 600 seconds of simulated gameplay (6\textasciitilde 20 game sets), we tracked various performance metrics, including game outcome, ball procession, the average number of passes, ball losses, and game duration. Sec. \ref{sec:adversaries} will provide additional details about the adversaries.  
Identical to the training setup, we spawn agents uniformly in their designated areas. While the spawning distribution is identical between the two teams in all scenarios, we biased the ball spawn location deep into the blue side for \textbf{Offensive}, into the red side for \textbf{Defensive} and kept them in the field center for \textbf{Equal} scenario. 

Additionally, we provide insight into the trainee's behavior directly with trajectory rollouts (Fig. \ref{fig:3v3_behavior} and \ref{fig:behavior_overview}) and value function of the base and ball position (Fig. \ref{fig:value_function}).

\subsection{Adversaries}
\label{sec:adversaries}
For the adversaries, we have the trainee itself (\textbf{RL}), a simple heuristic bot (\textbf{Bot}), and lastly the HELIOS agent as a \ac{soa} benchmark (\textbf{HELIOS}). The heuristic bot is a role-based scripted bot that assigns the closest agent with greedy soccer logic, i.e., approaching the ball and immediately kicking toward the goal once the ball becomes kickable. It also has a goalkeeper and defender logic. It is a simple yet effective adversary but far from sufficient as a benchmark. Proper benchmarking, however, is currently very difficult due to the lack of unified \ac{marl} soccer frameworks. Most existing \ac{marl} policies are embedded in other custom simulators, which lack interoperability. For our benchmark, we created an interface server using the \ac{rc} Simulation2D protocol \cite{simluation2d}, allowing us to compete against algorithms from other Simulation2D teams within our Isaac Gym environment. We chose HELIOS since it is currently the best-performing framework in the \ac{rc} Simulation2D, winning a total of 6 world championships in the past. To reduce the handicap introduced by the sim-to-sim gap, we replicated the movement and kick model inside Isaac Gym to the best of our ability.

\section{RESULTS AND ANALYSES}
        \label{sec:results}
        \begin{figure}
    \centering
    \includegraphics[width=0.48\textwidth]{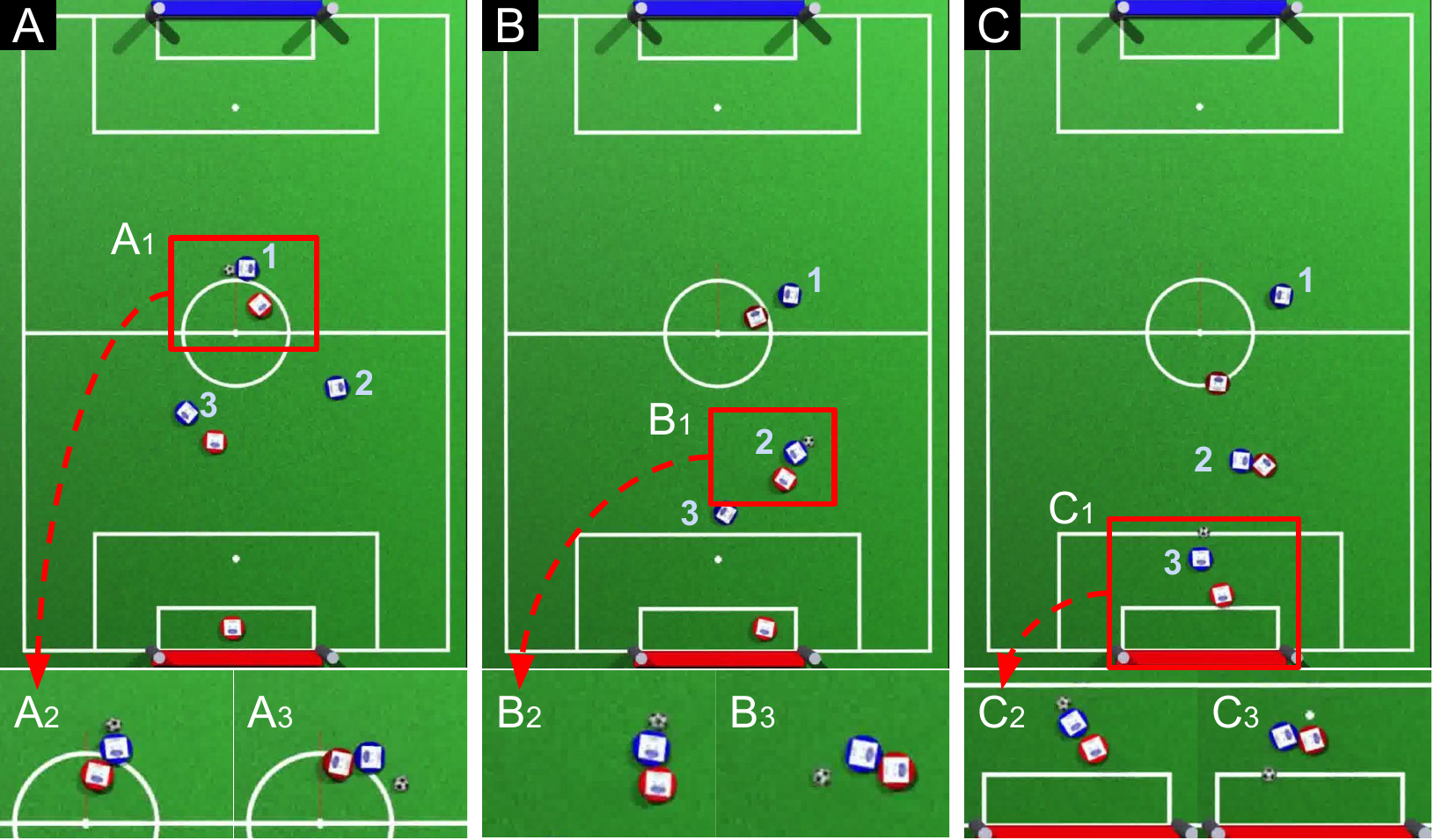}
    \caption{A top-down view sequence depicting \edited{some}{}typical trainee (blue) behaviors \edited{}{against bots} in a 3v3 game setting. A, B, and C provide an overview of the global behavior, such as positioning, while A1-3, B1-3, and C1-3 provide a zoom-in view showcasing numerous local ball-handling skills. }
    \label{fig:3v3_behavior}
\end{figure}

\begin{figure*}
        \centering
        \includegraphics[width=0.9\textwidth]{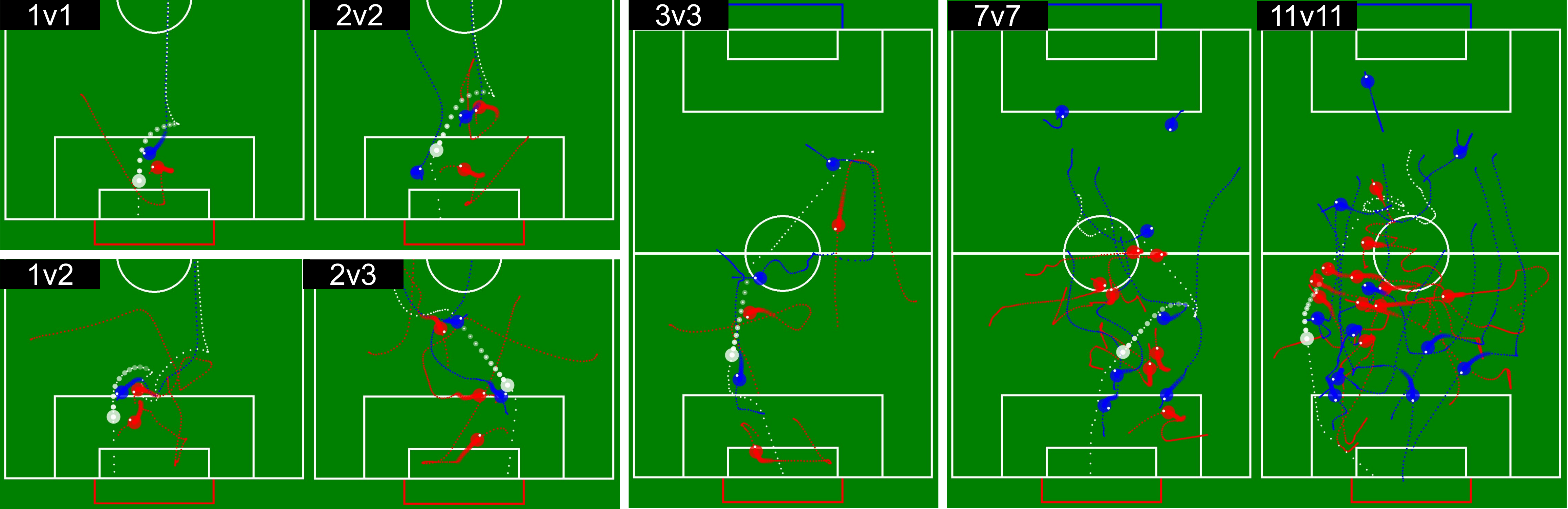}
        \caption{Recorded motion trajectories from various games illustrating generalized cooperative behavior for diverse team composition. The big colored dots represent the current position of our trainees (blue), adversaries (red), and the ball. The last few motion frames are additionally emphasized by the fading dot trails. The doted trajectories visualize the full game dynamic. The ball (white) trajectory, in particular, is useful for understanding how the ball is passed between trainees before they score the goal at the end. \edited{}{Note: we used bots for lower agent scenarios. The adversaries in the 7v7 and 11v11 games are directly controlled using the same RL policy.}}
        \label{fig:behavior_overview}
\end{figure*}

\subsection{Quantitative Results}
\label{sec:quantitative_results}
Fig. \ref{fig:eval_results} provides an overview of the evaluation result where the trainee policy performs against different adversary implementations. \edited{}{While typical team play emerges after 4000 epochs (\textasciitilde4h) of training . The evaluation results here were obtained from policies with 20k epochs (\textasciitilde21h) of training. }

Against itself (RL), all outcomes are roughly mirrored. Spawning the ball in one team's half directly leads to dominance, which can be seen from both game outcomes and ball ownership statistics. Once they have the ball, they perform numerous passes with a very high success rate.

Against the heuristic bot, the trainee excelled across all scenarios, especially in the offensive. When averaged over all scenarios, the trainees won about \SI{80}{\percent}, drew \SI{16}{\percent}, and lost in only \SI{4}{\percent} of all games and owning the ball most of the time. Although, more back and forth can be observed based on the pass and game duration statistics.

Finally, our trainees also prevailed against HELIOS by scoring more goals and delivering more consistent passes across all scenarios. From the significant ball ownership losses from both teams, we can conclude that both parties were actively engaging the ball. In total, our trainee won \SI{66.8}{\percent}, drew \SI{4.2}{\percent} and lost \SI{29}{\percent} of all games.

\edited{}{In addition to playing against fixed adversaries, we also conducted ablation studies to understand how dense rewards affect policy performance. We observed that the trainees failed to reach the ball without these dense rewards. Having permanent dense rewards allow the agent to learn decent gameplay, but suboptimal, winning only 12 and drawing 5 out of 100 games against the default policy.}

\subsection{Qualitative Results}
\label{sec:qualitative_results}
We depict \edited{some}{}common trainee behaviors in Fig. \ref{fig:3v3_behavior}, which provides a top-down view of a 3v3 game at different times. (A, B, and C) showcase global behavior and (A1-3, B1-3, C1-3) illustrate various local ball handling skills. Additionally, Fig. \ref{fig:behavior_overview} showcases \edited{some}{}common behaviors in games with other team compositions. Despite encountering only teams of up to 3v3 agents during training, the trainee has learned to generalize his behavior to a full-scale 11v11 game, showcasing various quality passes despite the clustering and chaos.  

Besides the policy behavior, Fig. \ref{fig:value_function} also depicts the trainees' value functions taken during a pass. It is worth noting that the ball position heat map (B1, B2) is nearly identical from both trainees' perspectives. This similarity aligns with our intuition since, unlike the base position, the ball's state is directly linked to the team's common objective, which is independent from the agent's perspective.

        
\section{CONCLUSION}
        \label{sec:conclusion}
        Our proposed approach allows effective end-to-end learning of cooperative behaviors in a multi-agent soccer environment and achieves an overall \SI{66.8}{\percent} win rate against the \ac{soa} benchmark HELIOS. By employing an improved \ac{gee} architecture, curricula, and self-play, the trainees can quickly explore and adopt intelligent behavior against previously unseen team composition (up to 11v11) and adversaries. Most of the sophisticated cooperative behavior emerges after the initial exploration, including the removal of dense rewards. With only the sparse rewards, the policy can demonstrate clear role assignment (Blue 1 in Fig. \ref{fig:3v3_behavior} A) and strategic positioning (Blue 2 and 3 in A, B). Our trainees have also acquired various ball-handling skills such as dribbling (A1-3), passing (A3, B3), and goal shooting (C3). Furthermore, the value function (Fig. \ref{fig:value_function}) also clearly illustrates many of the intended strategies, such as ball assignment and positioning (A1, A2), opportunities for a pass, and goal (prominent blue funnel in B1, B2). \edited{We invite the readers to check out the supplementary video for better understanding}{For a more detailed illustration, we invited the readers to check out the supplementary video\footnote{\url{https://www.youtube.com/watch?v=klETyDnWO2w}}.}

One limitation of our approach is corner case handling, for instance, when the ball is stuck at the border or when the opponents have the ball and are too far away. Our trainees often start drifting aimlessly in these scenarios, making them poor defenders.  \edited{Future research could focus on automating edge state detection, which can be utilized for the initial position curriculum to address performance issues in corner cases}{}. Furthermore, a higher agent number in our environment frequently results in an agent cluster, sometimes stalls the game entirely. As the ultimate goal is to deploy such policies onto a robotic system, we conduct preliminary tests \edited{of the policy}{}inside the NomadZ-NG framework \cite{nomadz}, which simulates soccer games for NAO robots. While the movement and global positioning exhibit comparable behavior, the kick maps poorly onto humanoid robots due to the lack of continuous ball handling. 

Future works should address these limitations. As our environment is already embedded in 3D, we can explore additional 3D actions such as high kicks or even directly replace our agents with fully articulated humanoid robots, possibly using our work as a hierarchical or pre-training component. Moreover, our work focuses on perfect information games where the state of all agents is available without delay. Handling additional complexity stemming from limited field of view, occlusion, and communication delay could also be an exciting research direction.






\footnotesize
\bibliographystyle{bibliography/IEEEtran}
\bibliography{bibliography/references_env, bibliography/references_rsl, bibliography/references_algorithm, bibliography/references_literature}

\end{document}